\documentclass[10pt,a4paper,onecolumn]{article}
\usepackage{marginnote}
\usepackage{graphicx}
\usepackage{xcolor}
\usepackage{authblk,etoolbox}
\usepackage{titlesec}
\usepackage{calc}
\usepackage{tikz}
\usepackage{hyperref}
\hypersetup{colorlinks,breaklinks=true,
            urlcolor=[rgb]{0.0, 0.5, 1.0},
            linkcolor=[rgb]{0.0, 0.5, 1.0},
            citecolor = blue}
\usepackage{caption}
\usepackage{tcolorbox}
\usepackage{amssymb,amsmath}
\usepackage{ifxetex,ifluatex}
\usepackage{seqsplit}
\usepackage{xstring}
\usepackage{natbib}
\usepackage[T1]{fontenc}
\usepackage[utf8]{inputenc}
\usepackage{cleveref}
\usepackage{float}
\usepackage{orcidlink}
\usepackage{xspace}
\usepackage{listings}
\usepackage{comment}
\usepackage{markdown}
\usepackage{tikz}
\usetikzlibrary{shapes,arrows,shadows,fit}
\usetikzlibrary{positioning}
\usetikzlibrary{bayesnet}

\let\origfigure\figure
\let\endorigfigure\endfigure
\renewenvironment{figure}[1][2] {
    \expandafter\origfigure\expandafter[H]
} {
    \endorigfigure
}

\let\textttOrig=\texttt
\def\texttt#1{\expandafter\textttOrig{\seqsplit{#1}}}
\renewcommand{\seqinsert}{\ifmmode
  \allowbreak
  \else\penalty6000\hspace{0pt plus 0.02em}\fi}

\usepackage{abstract}

\makeatletter
\let\href@Orig=\href
\def\href@Urllike#1#2{\href@Orig{#1}{\begingroup
    \def\Url@String{#2}\Url@FormatString
    \endgroup}}
\def\href@Notdoi#1#2{\def\tempa{#1}\def\tempb{#2}%
  \ifx\tempa\tempb\relax\href@Urllike{#1}{#2}\else
  \href@Orig{#1}{#2}\fi}
\def\href#1#2{%
  \IfBeginWith{#1}{https://doi.org}%
  {\href@Urllike{#1}{#2}}{\href@Notdoi{#1}{#2}}}
\makeatother

\newlength{\cslhangindent}
\newlength{\csllabelwidth}
 {
  \setlength{\parindent}{0pt}
  \ifodd #1 \everypar{\setlength{\hangindent}{\cslhangindent}}\ignorespaces\fi
  \ifnum #2 > 0
  \setlength{\parskip}{#2\baselineskip}
  \fi
 }%
 {}
\usepackage{calc}

\usepackage[top=3.5cm, bottom=3cm, right=1.5cm, left=1.0cm,
            headheight=2.2cm, reversemp, includemp, marginparwidth=4.5cm]{geometry}

\titleformat{\section}
  {\normalfont\sffamily\Large\bfseries}
  {}{0pt}{}
\titleformat{\subsection}
  {\normalfont\sffamily\large\bfseries}
  {}{0pt}{}
\titleformat{\subsubsection}
  {\normalfont\sffamily\bfseries}
  {}{0pt}{}
\titleformat*{\paragraph}
  {\sffamily\normalsize}

\usepackage{fancyhdr}
\makeatletter
\let\ps@plain\ps@fancy
\fancyheadoffset[L]{4.5cm}
\fancyfootoffset[L]{4.5cm}

\definecolor{linky}{rgb}{0.0, 0.5, 1.0}

\newtcolorbox{repobox}
   {colback=red, colframe=red!75!black,
     boxrule=0.5pt, arc=2pt, left=6pt, right=6pt, top=3pt, bottom=3pt}

\newcommand{\ExternalLink}{%
   \tikz[x=1.2ex, y=1.2ex, baseline=-0.05ex]{%
       \begin{scope}[x=1ex, y=1ex]
           \clip (-0.1,-0.1)
               --++ (-0, 1.2)
               --++ (0.6, 0)
               --++ (0, -0.6)
               --++ (0.6, 0)
               --++ (0, -1);
           \path[draw,
               line width = 0.5,
               rounded corners=0.5]
               (0,0) rectangle (1,1);
       \end{scope}
       \path[draw, line width = 0.5] (0.5, 0.5)
           -- (1, 1);
       \path[draw, line width = 0.5] (0.6, 1)
           -- (1, 1) -- (1, 0.6);
       }
   }

\patchcmd{\@maketitle}{center}{flushleft}{}{}
\patchcmd{\@maketitle}{center}{flushleft}{}{}
\patchcmd{\@maketitle}{\LARGE}{\LARGE\sffamily}{}{}

\def\maketitle{{%
  
  \g@addto@macro\AB@authors{%
    \protect\\[0.5em]%
    \protect\normalfont\protect\itshape on behalf of the Euclid Consortium%
  }%
  \AB@maketitle}}
  
\renewcommand\AB@affilsepx{ \protect\Affilfont}
\renewcommand\AB@affilnote[1]{{\bfseries #1}\hspace{3pt}}
\renewcommand{\affil}[2][]%
   {\newaffiltrue\let\AB@blk@and\AB@pand
      \if\relax#1\relax\def\AB@note{\AB@thenote}\else\def\AB@note{#1}%
        \setcounter{Maxaffil}{0}\fi
        \begingroup
        \let\href=\href@Orig
        \let\texttt=\textttOrig
        \let\protect\@unexpandable@protect
        \def\thanks{\protect\thanks}\def\footnote{\protect\footnote}%
        \@temptokena=\expandafter{\AB@authors}%
        {\def\\{\protect\\\protect\Affilfont}\xdef\AB@temp{#2}}%
         \xdef\AB@authors{\the\@temptokena\AB@las\AB@au@str
         \protect\\[\affilsep]\protect\Affilfont\AB@temp}%
         \gdef\AB@las{}\gdef\AB@au@str{}%
        {\def\\{, \ignorespaces}\xdef\AB@temp{#2}}%
        \@temptokena=\expandafter{\AB@affillist}%
        \xdef\AB@affillist{\the\@temptokena \AB@affilsep
          \AB@affilnote{\AB@note}\protect\Affilfont\AB@temp}%
      \endgroup
       \let\AB@affilsep\AB@affilsepx
}
\makeatother

\renewcommand\Affilfont{\sffamily\small\mdseries}
\ifnum 0\ifxetex 1\fi\ifluatex 1\fi=0 
  \usepackage[T1]{fontenc}
  \usepackage[utf8]{inputenc}

\else 
  \ifxetex
    \usepackage{mathspec}
    \usepackage{fontspec}

  \else
    \usepackage{fontspec}
  \fi
  \defaultfontfeatures{Ligatures=TeX,Scale=MatchLowercase}

\fi
\IfFileExists{upquote.sty}{\usepackage{upquote}}{}
\IfFileExists{microtype.sty}{%
\usepackage{microtype}
\UseMicrotypeSet[protrusion]{basicmath} 
}{}

\let\addcontentslineOrig=\addcontentsline
\def\addcontentsline#1#2#3{\bgroup
  \let\texttt=\textttOrig\addcontentslineOrig{#1}{#2}{#3}\egroup}
\let\markbothOrig\markboth
\def\markboth#1#2{\bgroup
  \let\texttt=\textttOrig\markbothOrig{#1}{#2}\egroup}
\let\markrightOrig\markright
\def\markright#1{\bgroup
  \let\texttt=\textttOrig\markrightOrig{#1}\egroup}

\usepackage{graphicx,grffile}
\makeatletter
\def\maxwidth{\ifdim\Gin@nat@width>\linewidth\linewidth\else\Gin@nat@width\fi}
\def\maxheight{\ifdim\Gin@nat@height>\textheight\textheight\else\Gin@nat@height\fi}
\makeatother
\setkeys{Gin}{width=\maxwidth,height=\maxheight,keepaspectratio}
\IfFileExists{parskip.sty}{%
\usepackage{parskip}
}{
\setlength{\parindent}{0pt}
\setlength{\parskip}{6pt plus 2pt minus 1pt}
}
\providecommand{\tightlist}{%
  \setlength{\itemsep}{0pt}\setlength{\parskip}{0pt}}
\ifx\paragraph\undefined\else
\let\oldparagraph\paragraph
\renewcommand{\paragraph}[1]{\oldparagraph{#1}\mbox{}}
\fi
\ifx\subparagraph\undefined\else
\let\oldsubparagraph\subparagraph
\renewcommand{\subparagraph}[1]{\oldsubparagraph{#1}\mbox{}}
\fi

\DeclareFixedFont{\ttb}{T1}{txtt}{bx}{n}{10} 
\DeclareFixedFont{\ttm}{T1}{txtt}{m}{n}{10}  

\usepackage{color}
\definecolor{deepblue}{rgb}{0,0,0.5}
\definecolor{deepred}{rgb}{0.6,0,0}
\definecolor{deepgreen}{rgb}{0,0.5,0}

\usepackage{listings}

\newcommand\pythonstyle{\lstset{
language=Python,
basicstyle=\scriptsize,
morekeywords={self},              
keywordstyle=\scriptsize\color{deepblue},
emph={MyClass,__init__},          
emphstyle=\ttb\color{deepred},    
stringstyle=\color{deepgreen},
    numbers=left,
frame=tb,                         
showstringspaces=false
}}

\lstnewenvironment{python}[1][]
{
\pythonstyle
\lstset{#1}
}
{}

\newcommand\pythoninline[1]{{\pythonstyle\lstinline!#1!}}

\newcommand{\pkgfont}{\texttt} 
\newcommand{\pkg}[2][]{%
  \if\relax\detokenize{#1}\relax
    \pkgfont{#2}%
  \else
    \href{#1}{\pkgfont{#2}}%
  \fi
}

\usepackage{listings}
\usepackage{xcolor}

\lstdefinelanguage{YAML}{
  keywords={true, false, null},
  keywordstyle=\color{blue}\bfseries,
  comment=[l]{\#},
  commentstyle=\color{gray}\itshape,
  stringstyle=\color{teal},
  morestring=[b]",
  morestring=[b]',
}

\newcommand*{\AckInstitutions}{a number of agencies and
  institutes that have supported the development of \textit{Euclid}, in
  particular
  the Agenzia Spaziale Italiana,
  the Austrian Forschungsf\"orderungsgesellschaft funded through BMIMI,
  the Belgian Science Policy,
  the Canadian Euclid Consortium,
  the Deutsches Zentrum f\"ur Luft- und Raumfahrt,
  the DTU Space and the Niels Bohr Institute in Denmark,
  the French Centre National d'Etudes Spatiales,
  the Funda\c{c}\~{a}o para a Ci\^{e}ncia e a Tecnologia,
  the Hungarian Academy of Sciences,
  the Ministerio de Ciencia, Innovaci\'{o}n y Universidades,
  the National Aeronautics and Space Administration,
  the National Astronomical Observatory of Japan,
  the Netherlandse Onderzoekschool Voor Astronomie,
  the Norwegian Space Agency,
  the Research Council of Finland,
  the Romanian Space Agency,
  the Swiss Space Office (SSO) at the State Secretariat for Education, Research, and Innovation (SERI),
  and the United Kingdom Space Agency.
  A complete and detailed list is available on the \textit{Euclid}\ web site
  (\url{www.euclid-ec.org/consortium/community/}).\xspace}
  
\newcommand{\AckEC}{The Euclid Consortium acknowledges the European
  Space Agency and \AckInstitutions}

\title{\texttt{cloelib}: A Flexible Python Library for Computing Cosmological
Observables in the Euclid Era}

\date{\vspace{-7ex}}

\begin{document}
\author[1]{(cloe-org maintainers:) Marco Bonici\orcidlink{0000-0002-8430-126X}}
\author[2]{Guadalupe Cañas-Herrera\thanks{Corresponding author: canasherrera@strw.leidenuniv.nl}\orcidlink{0000-0003-2796-2149}}
\author[3]{Pedro Carrilho\orcidlink{0000-0003-1339-0194}}
\author[4]{Santiago Casas\orcidlink{0000-0002-4751-5138}}
\author[5]{Chiara Moretti\orcidlink{0000-0003-3314-8936}}
\author[6]{Andrea Pezzotta\orcidlink{0000-0003-0726-2268}}
\author[5]{(cloelib contributors:) Michel Aguena\orcidlink{0000-0001-5679-6747}}
\author[7]{Giovanni Arico\orcidlink{0000-0002-2802-2928}}
\author[8]{Zahra Baghkhani\orcidlink{0000-0002-6632-2614}}
\author[9,10]{Matteo Baratto\orcidlink{0009-0000-8702-9591}}
\author[11]{Emilio Bellini\orcidlink{0000-0003-4762-0795}}
\author[2]{Jip de Buck\orcidlink{0009-0001-5175-9282}}
\author[12]{Klara Bertmann\orcidlink{0009-0004-6700-2470}}
\author[13,14]{Ben Bose\orcidlink{0000-0003-1965-8614}}
\author[2,15]{Jeger C. Broxterman\orcidlink{0000-0002-8155-5977}}
\author[1]{Pierre Burger\orcidlink{0000-0002-6374-5208}}
\author[10]{Carmelita Carbone\orcidlink{0000-0003-0125-3563}}
\author[16]{Chaitanya Chawak}
\author[8]{Jose Coloma-Nadal\orcidlink{0009-0003-0538-4349}}
\author[8]{Martin Crocce\orcidlink{0000-0002-9745-6228}}
\author[17]{Stefano Davini\orcidlink{0000-0003-3269-1718}}
\author[13]{Christopher A. J. Duncan}
\author[10]{Samuel Farrens\orcidlink{0000-0002-9594-9387}}
\author[13,14]{Lisa Goh\orcidlink{0000-0002-0104-8132}}
\author[18]{Nastassia Grimm\orcidlink{0000-0001-9602-0599}}
\author[13]{Alex Hall\orcidlink{0000-0002-3139-8651}}
\author[10,19]{Ryusei R. Kano\orcidlink{0009-0002-9108-8396}}
\author[20]{Martin Kärcher\orcidlink{0000-0001-5868-647X}}
\author[21]{Felicitas Keil\orcidlink{0000-0002-8108-1679}}
\author[22]{Raphaël Kou\orcidlink{0000-0003-3408-3062}}
\author[23]{Laila Linke\orcidlink{0000-0002-2622-8113}}
\author[24,25]{Arthur Loureiro\orcidlink{0000-0002-4371-0876}}
\author[26]{Katarina Markovic\orcidlink{0000-0001-6764-073X}}
\author[2]{David Navarro-Gironés\orcidlink{0000-0003-0507-372X}}
\author[17]{Filippo Oppizzi\orcidlink{0000-0003-3904-8370}}
\author[8]{Gabriele Parimbelli\orcidlink{0000-0002-2539-2472}}
\author[27]{Anna Porredon\orcidlink{0000-0002-2762-2024}}
\author[28]{Robert Reischke\orcidlink{0000-0001-5404-8753}}
\author[29]{Jaime Ruiz Zapatero\orcidlink{0000-0002-7951-4391}}
\author[6]{Iñigo Sáez-Casares\orcidlink{0000-0003-0013-5266}}
\author[30]{Ziad Sakr\orcidlink{0000-0002-4823-3757}}
\author[31]{Neel Shah\orcidlink{0009-0001-4424-6489}}
\author[32]{Davide Sciotti\orcidlink{0009-0008-4519-2620}}
\author[15,2]{Matthieu Schaller\orcidlink{0000-0002-2395-4902}}
\author[33]{Ivan Sladoljev\orcidlink{0009-0002-9702-2101}}
\author[27]{Arghavan Souki\orcidlink{0009-0000-4771-7728}}
\author[34]{Sankarshana Srinivasan \orcidlink{0000-0003-1539-3276}}
\author[35]{Konstantinos Tanidis\orcidlink{0000-0001-9843-5130}}
\author[36]{Peter L. Taylor\orcidlink{0000-0001-6999-4718}}
\author[37]{Nicolas Tessore\orcidlink{0000-0002-9696-7931}}
\author[13,14]{Linus Thummel\orcidlink{0000-0002-9807-5494}}
\author[13,14]{Maria Tsedrik\orcidlink{0000-0002-0020-5343}}
\author[8,38,21]{Isaac Tutusaus\orcidlink{0000-0002-3199-0399}}
\author[2]{Casper Vedder\orcidlink{0009-0007-6341-4648}}
\author[12]{Angus H. Wright\orcidlink{0000-0001-7363-7932}}
\author[39]{Miguel Zumalacarregui\orcidlink{0000-0002-9943-6490}}
\author[13]{Joe Zuntz\orcidlink{0000-0001-9789-9646}}
\author[ ]{, on behalf of the Euclid Consortium}

\affil[1]{Waterloo Centre for Astrophysics, University of Waterloo, Canada}
\affil[2]{Leiden Observatory, Leiden University, Netherlands}
\affil[3]{Centre for Astrophysics Research, University of Hertfordshire, United Kingdom}
\affil[4]{Scientific Information, German Aerospace Center, Germany}
\affil[5]{INAF - Osservatorio Astronomico di Trieste, Italy}
\affil[6]{INAF - Osservatorio Astronomico di Brera, Italy}
\affil[7]{INFN - Sezione di Bologna, Italy}
\affil[8]{Institute of Space Sciences, Spain}
\affil[9]{Department of Physics, Università degli Studi di Milano, Italy}
\affil[10]{INAF - IASF Milano, Italy}
\affil[11]{INFN - Sezione di Trieste, Italy}
\affil[12]{Astronomical Institute, Ruhr University Bochum, Germany}
\affil[13]{Institute for Astronomy, University of Edinburgh, United Kingdom}
\affil[14]{Higgs Centre for Theoretical Physics, University of Edinburgh, United Kingdom}
\affil[15]{Lorentz Institute for Theoretical Physics, Leiden University, Netherlands}
\affil[16]{CEA Paris-Saclay, France}
\affil[17]{INFN - Sezione di Genova, Italy}
\affil[18]{Department of Physics, University of Oxford, United Kingdom}
\affil[19]{Division of Particle and Astrophysical Science, Nagoya University, Japan}
\affil[20]{Dipartimento di Fisica "Aldo Pontremoli", Università degli Studi di Milano, Italy}
\affil[21]{IRAP, Université de Toulouse, France}
\affil[22]{Department of Physics \& Astronomy, University of Sussex, United Kingdom}
\affil[23]{Institut für Astro- und Teilchenphysik, Universität Innsbruck, Austria}
\affil[24]{Oskar Klein Centre for Cosmoparticle Physics, Department of Physics, Stockholm University, Stockholm, SE-106 91, Sweden}
\affil[25]{Astrophysics Group, Blackett Laboratory, Imperial College London, London SW7 2AZ, UK}
\affil[26]{Jet Propulsion Laboratory, USA}
\affil[27]{CIEMAT, Spain}
\affil[28]{Argelander-Institut für Astronomie, Universität Bonn, Germany}
\affil[29]{Advanced Research Computing Centre, University College London, United Kingdom}
\affil[30]{IFT, Spain}
\affil[31]{University of Portsmouth, United Kingdom}
\affil[32]{Osservatorio Astronomico di Roma, Italy}
\affil[33]{Department of Physics, Royal Holloway, University of London, United Kingdom}
\affil[34]{Universitat Sternwarte, Ludwig Maximilian Universitat, Germany}
\affil[35]{Center for Astrophysics and Cosmology, University of Nova Gorica, Slovenia}
\affil[36]{CCAPP, The Ohio State University, USA}
\affil[37]{Mullard Space Science Laboratory, University College London, United Kingdom}
\affil[38]{Institut d'Estudis Espacials de Catalunya (IEEC), Spain}
\affil[39]{Max Planck Institute for Gravitational Physics, Germany}

\maketitle

\marginpar{
  \begin{flushleft}
  \vspace{-500pt}
  \sffamily\small

  {\bfseries DOI: N/A}

  \vspace{2mm}

  {\bfseries Software}
  \begin{itemize}
    \setlength\itemsep{0em}
    \item \href{N/A}{\color{linky}{Review}} \ExternalLink
    \item \href{https://github.com/cloe-org/cloelib}{\color{linky}{Repository}} \ExternalLink
    \item \href{https://zenodo.org/record/TBD}{\color{linky}{Archive (Zenodo)}} \ExternalLink
  \end{itemize}

  \vspace{2mm}
  \par\noindent\hrulefill\par
  \vspace{2mm}

  {\bfseries Editor:} \href{https://example.com}{Pending editor} \ExternalLink \\
  \vspace{1mm}
  {\bfseries Reviewers:}
  \begin{itemize}
    \setlength\itemsep{0em}
    \item \href{https://github.com/pending}{@pending}
  \end{itemize}

  \vspace{2mm}
  {\bfseries Submitted:} N/A \\
  {\bfseries Published:} N/A

  \vspace{2mm}
  {\bfseries License}\\
  Authors retain copyright and release the work under CC BY 4.0
  (\href{http://creativecommons.org/licenses/by/4.0/}{\color{linky}{link}}).
  \end{flushleft}
}

\vspace{.2cm}

\vspace{.5cm}

\section{Summary}
\texttt{cloelib},
\href{https://github.com/cloe-org/cloelib}{cloe-org/cloelib}, is a
Python library developed to compute cosmological observables within the
Cosmology Likelihood for Observables in Euclid (\texttt{CLOE}) project
\href{https://github.com/cloe-org}{\textbf{cloe-org}}\footnote{\href{https://github.com/cloe-org}{https://github.com/cloe-org}}.
As cosmology enters a precision era driven by galaxy survey missions
such as \emph{Euclid}, there is a growing need for flexible, efficient,
and differentiable software capable of supporting next-generation
inference pipelines. \texttt{cloelib} addresses these demands through a
modular architecture that interfaces seamlessly with established
Boltzmann solvers whilst incorporating JAX-based automatic
differentiation to enable gradient-based methods. The library defines
consistent protocols for background evolution, perturbations, and
non-linear structure formation, and supports a wide range of
observables, including photometric and spectroscopic large-scale
structure probes, as well as cross-correlations with the Cosmic
Microwave Background and galaxy clusters. In its finalised form,
\texttt{cloelib} is intended to serve as the reference theory
computation infrastructure for Euclid's first cosmological release,
bridging traditional numerical cosmology with modern optimisation
techniques and emerging machine learning approaches to inference.

\hypertarget{statement-of-need}{%
\section{Statement of need}\label{statement-of-need}}

The field of observational cosmology is undergoing a rapid
transformation, driven by the advent of Stage IV galaxy surveys
such as the European Space Agency's \emph{Euclid} mission \citep{Euclid:2024}, the Dark Energy Spectroscopic
Instrument (DESI; \citealt{DESI_review}), the
\textit{Vera C. Rubin} Observatory's Large Synoptic Survey Telescope
\citep{LSST}, and NASA's \textit{Nancy Grace Roman} Space
Telescope\footnote{\href{https://roman.gsfc.nasa.gov/science/ccs/ROTAC-Report-20250424-v1.pdf}{https://roman.gsfc.nasa.gov/science/ccs/ROTAC-Report-20250424-v1.pdf}}.
These projects are generating vast volumes of high-quality data, mapping
the large-scale structure of the Universe with unprecedented precision.
Extracting robust scientific insights from these data requires the efficient computation of theoretical predictions that can be directly and reliably compared with observations to constrain cosmological models. This poses stringent demands on computational tools, which must accurately capture complex theoretical scenarios while remaining computationally efficient.
Despite significant progress, existing cosmological software frameworks often lack the flexibility needed to seamlessly integrate diverse, pre-existing components and to explore a broad range of theoretical models alongside comprehensive treatments of systematic effects. \texttt{cloelib} addresses this gap by providing a highly flexible and extensible platform for computing theoretical predictions across multiple large-scale structure observables under a wide variety of cosmological models. In doing so, it enables faster and more streamlined statistical analyses, helping to meet the demands of the next generation of precision cosmology experiments.

In this context, \texttt{cloelib} is a fully Pythonic library for observational modelling, designed to operationalise the flexibility and efficiency required for modern cosmological inference. It represents a natural evolution of the structural formalism originally developed in the Cosmology Likelihood for Observables in Euclid (\texttt{CLOE}) software, designed by the Euclid Consortium, extending it towards more advanced use cases and significantly enhanced capabilities beyond those presented in \citet{EP-CLOE2}. The original \texttt{CLOE}\footnote{\href{https://github.com/cloe-org/CLOE}{https://github.com/cloe-org/CLOE}} software has played a central role in numerous Euclid analyses—see \citet{Euclid:2024}, \citet{EP-CLOE3}, \citet{EP-CLOE4}, \citet{EP-CLOE5} and \citet{EP-CLOE6}—demonstrating its robustness and scientific impact in forecasting and validating \emph{Euclid} performance.
However, the increasing complexity of future cosmological analyses—such as the joint treatment of multiple probes, the inclusion of high-dimensional nuisance parameter spaces, and the combination of heterogeneous datasets—has exposed structural limitations in its original design. In practice, extending \texttt{CLOE} to accommodate new observables or modelling choices often required intrusive modifications across multiple parts of the software, leading to the accumulation of technical debt and reduced maintainability over time. Moreover, the framework was not originally conceived to support the independent development and seamless integration of new theoretical models, systematics, or data components within a unified pipeline.
As a result, a substantial restructuring became necessary to meet these emerging requirements. \texttt{cloelib} builds directly on the conceptual and practical foundations laid by \texttt{CLOE}, whilst introducing a redesigned architecture in which components—such as theory predictions—are decoupled and interact through well-defined interfaces. This enables flexible composition of analysis pipelines, facilitates the inclusion of new physics or datasets, and improves scalability for large parameter spaces.

\section{State of the Field}
\label{sec:state_of_field}

Similarly to \texttt{CCL} \citep{pyccl}, \texttt{CosmoSIS} \citep{CosmoSIS}, \texttt{CAMB} \citep{Lewis:2000}, \texttt{CLASS} \citep{Blas:2011}, CosmoLike \citep{CosmoLike}, and \texttt{CoCoA}\footnote{\href{https://github.com/CosmoLike/cocoa}{https://github.com/CosmoLike/cocoa}}, it supports the computation of large-scale structure probes in the form of angular or spatial two-point correlations, and associated observables such as cosmic shear (including cross-correlations with the cosmic microwave background), galaxy clustering, and spectroscopic power spectrum multipoles.
Yet, \texttt{cloelib} is the first and only large-scale structure code in the cosmology community to implement a unified interface to multiple cosmological codes using Python protocols. This design enables researchers to seamlessly switch between different theoretical implementations (backends), such as Boltzmann solvers or emulators, without modifying their analysis pipelines or the internal workings of \texttt{cloelib}. As a result, new theoretical models or external codes can be incorporated with minimal changes, and different implementations can be compared within a consistent analysis setup. This capability naturally supports the integration of additional pipelines and promotes rapid experimentation in cosmological analyses. In fact, following this protocol-based framework, \texttt{cloelib} already interfaces with several well-established cosmological codes in the community.

Among these tools, for the computation of background quantities and the
matter power spectrum, \texttt{cloelib} provides interfaces to widely
used Boltzmann solvers such as \texttt{CAMB} and \texttt{CLASS}, as well
as their extensions [e.g.~\texttt{hi\_class} \citep{hi_class_1, hi_class_2}, \texttt{mgclass\ II} \citep{Sakr_2022}, and \texttt{mochi\_class} \citep{mochi_class}]. To accelerate matter power spectrum evaluations, \texttt{cloelib} also
supports a range of state-of-the-art emulators, including
\texttt{CosmoPower} \citep{SpurioMancini:2021, Piras23}, \texttt{BACCOemu} \citep{bacco-full-power, bacco-original, bacco-tracers, bacco-euclid, bacco-emu-baryons},
\texttt{EuclidEmulator2}\footnote{\href{https://github.com/PedroCarrilho/EuclidEmulator2/tree/pywrapper}{https://github.com/PedroCarrilho/EuclidEmulator2}}
\citep{EE2}, and \texttt{HMCode2020Emu} \citep{Mead:2021, Tsedrik2024}, as
well as nonlinear model extensions such as \texttt{ReACT} \citep{ReACT} and baryonic effects through the \texttt{FlamingoBaryonResponseEmulator} \citep{flamingo-emu}. These emulators provide orders-of-magnitude speed-ups while
retaining percent-level accuracy, making them well suited for modern
cosmological inference pipelines. To compute spectroscopic observables, \texttt{cloelib} interfaces with the state-of-the-art theory code
\texttt{PBJ}\footnote{\href{https://gitlab.com/chiaramoretti/pbj}{https://gitlab.com/chiaramoretti/pbj}}\citep{Moretti:2023, pbj} and the emulator \texttt{comet-emu} \citep{Eggemeier:2022, Pezzotta:2025} for fast computation of nonlinear spectroscopic galaxy clustering. This modular design, combined with its
computational performance, makes \texttt{cloelib} particularly well suited for systematic studies, model comparison, and robust cross-validation of cosmological results.

A key feature of \texttt{cloelib} is its native integration with JAX \citep{jax2018github}, enabling automatic differentiation for
cosmological observables, efficient gradient-based computation, and
advanced just-in-time (\texttt{jit}) compilation. This transforms
conventional cosmological pipelines into fully differentiable
programmes, making advanced inference techniques---such as Hamiltonian
Monte Carlo and neural network-based modelling---readily accessible.
Whilst such methods are often difficult to implement efficiently in
traditional frameworks, \texttt{cloelib} is designed to facilitate these
workflows, offering a robust and flexible platform for developing neural
network emulators and exploring new inference methodologies.

In addition, \texttt{cloelib} serves the practical needs of both the
Euclid Collaboration and the wider cosmology community by offering
implementations of survey-specific systematics, such as
Alcock--Paczynski corrections, shear and photometric redshift
calibration parameters, and spectroscopic purity in surveys. The library
works seamlessly with \texttt{cloelike}, its companion likelihood
module, which supports the computation of likelihoods for Euclid
observables such as cosmic shear, 2x2-pt, and 3x2-pt photometric
analyses, spectroscopic galaxy clustering and BAO, as well as their
combinations. Together, these tools enable end-to-end cosmological
analyses, covering the full chain from observable computation to
likelihood evaluation and posterior sampling for parameter inference.

Beyond its scientific scope, \texttt{cloelib} is designed for efficiency
and consistency in cosmological applications, featuring native source
code implementations of theoretical predictions and \texttt{jit} caching
mechanisms that accelerate the computation of otherwise expensive
integrals. The library is structured to support portability and
reproducibility across different environments, facilitating tasks such
as running inference chains and integrating with broader analysis
pipelines. It also incorporates testing infrastructure and performance
profiling tools to ensure robustness and scalability. By combining
theoretical flexibility, computational performance, and modern
programming practices within an open science framework, \texttt{cloelib}
contributes to the computational toolkit for precision cosmology and is
well suited for large-scale structure analyses in the coming decade.

\section{Software Design}
\label{sec:software}
The architecture of \texttt{cloelib} is built around a clear separation of concerns to be able to seamlessly compute theoretical prediction of cosmological observables. It allows to wrap up theoretical predictions around common cosmological frameworks such as \texttt{Cobaya} \citep{Cobaya}. \texttt{cloelib} organises functionality into four distinct layers: cosmological backgrounds (e.g., expansion history and distances), perturbation theory (e.g., linear and non-linear matter power spectra), observables (e.g., cosmic shear and photometric galaxy clustering, spectroscopic clustering full shape and post-reconstruction BAO), and summary statistics (e.g., angular power spectra and correlation functions). This layered design allows researchers to flexibly mix and match different theoretical models and numerical approximations, supporting both standard analyses and experimental workflows. For example, users can compute angular power spectra using the Limber approximation with any combination of supported Boltzmann solvers and non-linear models, or define custom window functions for specific survey geometries.

Each layer is defined by a Python protocol (PEP 544), enforcing a "plug-and-play" approach to modularity. Concretely, the library is organised into specialised modules: cosmology backends implementing the \texttt{Background} and \texttt{Perturbations} protocols, observable modules providing window functions and power spectrum interfaces through the \texttt{Tracer} and \texttt{SpectroPower} protocols, summary statistics for angular correlations and Legendre multipoles, and auxiliary utilities for mathematical operations and caching. The \texttt{Perturbations} protocol supports Python mixins to enable modular class composition, allowing users to extend or modify matter power spectrum functionality—such as adding baryonic effects—without altering the underlying implementation.

Performance-critical sections utilise \texttt{jit} compilation, while the caching system avoids redundant evaluations across repeated calculations. In this sense, \texttt{Background}, \texttt{Perturbations}, \texttt{Tracer}, and \texttt{SpectroPower} are structural interfaces that guarantee type safety and extensibility without relying on inheritance hierarchies. Users can include only the components they need, choose among interchangeable backends, and combine them freely—all without altering the core logic of their pipeline. This architecture ensures robustness, reusability, and ease of experimentation by design.

The library integrates with the broader Python scientific ecosystem through \texttt{NumPy} and \texttt{SciPy}, while maintaining full compatibility with JAX arrays for differentiable computations. This enables researchers to construct complex, end-to-end analysis pipelines that are simultaneously computationally efficient, maintainable, and—where needed—fully differentiable.





\section{Usage Examples}\label{usage-examples}
The power of \texttt{cloelib} lies in its intuitive API that allows
researchers to quickly set up complex cosmological calculations using
this ``plug-and-play'' approach. Here we demonstrate key features
through practical examples.


\hypertarget{initializing-cosmological-models}{%
\subsection{Initializing Cosmological
Models}\label{initializing-cosmological-models}}

\texttt{cloelib} provides a consistent interface for different
cosmological backends. Users can instantiate cosmological models using
standard parameters. The example below demonstrates how different codes can be used interchangeably, allowing for straightforward cross-validation of results:
\begin{lstlisting}[language=python]
from cloelib.cosmology.camb_cosmology import (
    CAMBBackground,
)
from cloelib.cosmology.jax_cosmology import (
    JAXBackground,
)
import numpy as np
# --- Cosmological parameters ---
H0 = 70.0
Omega_cdm0 = 0.25
Omega_b0 = 0.05
As = 2e-9
ns = 0.96
w0 = -1.0
wa = 0.0
Omega_k0 = 0.0
mnu = 0.06
N_mnu = 1
gamma_MG = 0.545
# --- Initialize background objects ---
camb_bg = CAMBBackground(
    H0=H0, Omega_cdm0=Omega_cdm0, Omega_b0=Omega_b0,
    As=As, ns=ns, w0=w0, wa=wa, Omega_k0=Omega_k0,
    mnu=mnu, N_mnu=N_mnu, gamma_MG=gamma_MG
)
jax_bg = JAXBackground(
    H0=H0, Omega_cdm0=Omega_cdm0, Omega_b0=Omega_b0,
    As=As, ns=ns, w0=w0, wa=wa, Omega_k0=Omega_k0,
    mnu=mnu, N_mnu=N_mnu, gamma_MG=gamma_MG
)
# --- Compute background quantities ---
z = np.linspace(0, 3, 256)
H_z_camb = camb_bg.hubble_parameter(z)
H_z__jax = jax_bg.hubble_parameter(z)
\end{lstlisting}

\hypertarget{Computing-Power-Spectra}{\subsection{Computing Power Spectra}\label{Computing-Power-Spectra}}

The library supports both linear and non-linear perturbation theories:

\begin{lstlisting}[language=python]
# Initialize perturbations

from cloelib.cosmology.camb_cosmology import (
    CAMBLinearPerturbations,
)
from cloelib.cosmology.jax_cosmology import (
    JAXNonLinearPerturbations,
)
from cloelib.cosmology.HMcode2020Emu_cosmology import (
    HMemuLinearPerturbations, HMemuNonLinearPerturbations,
)

# Linear perturbations
camb_linear = CAMBLinearPerturbations(background=camb_bg, 
                                      redshifts=z)
hmcode2020_linear = HMemuLinearPerturbations(background=camb_bg, redshifts=z)

# Non-linear perturbations
jax_nonlinear = JAXNonLinearPerturbations(background=jax_bg)
hmcode2020emu_nonlinear = HMemuNonLinearPerturbations(background=camb_bg, linearperturbations=hmcode2020_linear, redshifts=z, log10TAGN=7.8)

# Compute matter power spectra
ks = np.logspace(-4, np.log10(5), 256)
linear_pk_camb = camb_linear.matter_power_spectrum(z, ks)
linear_pk_hmcode2020 = hmcode2020_linear.matter_power_spectrum(z, ks)
nonlinear_pk_jax = jax_nonlinear.matter_power_spectrum(z, ks)
nonlinear_pk_hmcode2020emu = hmcode2020emu_nonlinear.matter_power_spectrum(z, ks)
\end{lstlisting}

\hypertarget{Photometric-Observables}{\subsection{Photometric Observables}\label{Photometric-Observables}}

\texttt{cloelib} provides robust computation of observables for
photometric surveys, including galaxy clustering, cosmic shear, and cosmic microwave background (CMB) lensing, with advanced modelling of systematic effects. Specifically, the calculation
of shear and position window functions is managed within the
\texttt{photo} module. Each tracer must be initialised with a
\texttt{Perturbations}-compatible instance, the galaxy density
distribution in redshift bins, and the relevant systematic models, such
as intrinsic alignments, galaxy bias, magnification, shear
multiplicative bias calibration nuisance parameters, and photometric
calibration nuisance parameters. Every computed window function is
accessible from each tracer instance. Tracers are combined into
two-point summary statistics, both in real and harmonic space, using the
functions available in the \texttt{summary\_statistics} module.

\begin{lstlisting}[language=python]
from cloelib.observables.photo import ShearTracer, PositionsTracer
from cloelib.summary_statistics.angular_two_point import AngularTwoPoint

# Define galaxy redshift distributions (normalized)
#my_dndz = my_dndz  # Shape: (n_bins, n_z_points)

# Create tracers with survey-specific nuisance parameters
# PositionsTracer requires per-bin photo-z shifts 
# and magnification bias
pos_nuisance = {
    'b1_photo_poly0': 1.2,
    **{f'b1_photo_poly{i}': 0.0 for i in range(1, 4)},
    **{f'magnification_bias_{i}': 0.0 for i in range(1, 7)},
    **{f'dz_pos_{i}': 0.0 for i in range(1, 7)},
    **{f'width_pos_{i}': 1.0 for i in range(1, 7)},
}
# Compact example for a 6-bin tomographic setup.
# Parameters are generated programmatically here, but can
# also be defined individually for full survey-specific control.
tracer_pos = PositionsTracer(
    perturbations=hmcode2020emu_nonlinear,
    dndz=my_dndz_pos_norm,
    z=zs,
    galaxy_bias_model='poly',
    nuisance_params=pos_nuisance
    }
)
# ShearTracer requires intrinsic alignment (IA) 
# and photo-z shift parameters
shear_nuisance = {
    'AIA': 1.72,
    'CIA': 0.0134,
    'EtaIA': -0.41,
    **{f'multiplicative_bias_{i}': 0.0 for i in range(1, 7)},
    **{f'dz_shear_{i}': 0.0 for i in range(1, 7)},
    **{f'width_shear_{i}': 1.0 for i in range(1, 7)},
}
tracer_she = ShearTracer(
    perturbations=hmcode2020emu_nonlinear,
    dndz=my_dndz_pos_norm,
    z=zs,
    nuisance_params=shear_nuisance
    }
)
# Compute angular power spectra using the Limber approximation
twopoint_posshe = AngularTwoPoint(tracer_pos, tracer_she)
twopoint_pospos = AngularTwoPoint(tracer_pos, tracer_pos)
twopoint_sheshe = AngularTwoPoint(tracer_she, tracer_she)


ells = np.arange(2, 3000)
cls_sheshe = twopoint_sheshe.get_Cl(ells, nl=0, ks=ks)
cls_posshe = twopoint_posshe.get_Cl(ells, nl=0, ks=ks)
cls_pospos = twopoint_pospos.get_Cl(ells, nl=0, ks=ks)
cls = {**cls_sheshe, **cls_posshe, **cls_pospos}
\end{lstlisting}

\hypertarget{Spectroscopic-Observables}{\subsection{Spectroscopic Observables}\label{Spectroscopic Observables}}

Calculation of the Legendre multipoles (both in Fourier and
configuration space) is handled by the \texttt{LegendreMultipoles}
module, which interfaces with objects that comply with the
\texttt{SpectroPower} protocol. This module implements shared modelling
layers that are handled coherently by cloelib, rather than relying on
individual implementations of external pipelines. Modelled effects
include shot-noise corrections, Alcock--Paczynski distortions, and the
convolution with the survey window function, as well as a number of
observational systematic effects, such as spectroscopic redshift errors
and the presence of contaminants. In addition, this module can compute
the two-point correlation function and projects it -- or \(P(k,\mu)\) --
to Legendre multipoles. As an example, we show below how to obtain a
prediction for the power spectrum multipoles using the
\texttt{comet-emu} package.

\begin{lstlisting}[language=python]
from cloelib.observables.CometEFT_spectro import CometEFT_SpectroPower
from cloelib.summary_statistics.legendre_multipoles import LegendreMultipoles
import numpy as np

# EFT bias and nuisance parameters for a single redshift bin
RSD_parameters = { # Bias and EFT counterterms
    'b1': 1.8, 'b2': 0.0, 'bG2': 0.0, 'bGam3': 0.0,
    'c0': 0.0, 'c2': 0.0, 'c4': 0.0
}

# Spectroscopic power spectrum at a single effective redshift
spectro_power = CometEFT_SpectroPower(
    background=camb_bg,
    RSD_parameters=RSD_parameters,
    redshift=1.0,
)

# Compute Legendre multipoles with Alcock-Paczynski corrections
noise_systematics_parameters = {
    'NP0': 1.0, 'NP20': 0.0, 'NP22': 0.0,# Shot-noise parameters
    'sigmaz': 0.0, # Redshift error for AP effect
    'fout': 0.0 # Purity
}
nbar = 1e-3  # galaxy number density [h/Mpc]^3
multipoles = LegendreMultipoles(
    spectro_power=spectro_power,
    background_fiducial=camb_bg,
    parameters=noise_systematics_parameters,
    nbar=nbar,
)

k = np.logspace(-2, np.log10(0.5), 80)
Pk_ell = multipoles.power_multipoles(k, ells=np.array([0, 2, 4]))
# Pk_ell is a dict: {'ell0': array, 'ell2': array, 'ell4': array}
\end{lstlisting}

The same interface is used to compute two-point correlation function multipoles via an FFTLog transform, and to apply convolutions with survey window functions.

\hypertarget{Protocol-Compliance-of-Interfaces}{\subsection{Protocol Compliance of Interfaces}\label{Protocol-Compliance-of-Interfaces}}

\texttt{cloelib} natively supports Python structural subtyping (PEP
544); the \texttt{Background} and \texttt{Perturbations} protocols are
marked with \texttt{@runtime\_checkable}, allowing explicit compliance
checks at the beginning of an analysis. New protocol interfaces can either be checked by the user as well as by continuous integration in the \texttt{GitHub} repository.

\begin{lstlisting}[language=python]
from cloelib.cosmology.cosmology import (
    Background,
)
# Protocol compliance is verified at runtime
assert isinstance(camb_bg, Background)
if isinstance(camb_bg, Background):
    print("CAMB background is compliant with the Background protocol.")
\end{lstlisting}
Because both objects conform to the same protocol, any downstream
\texttt{cloelib} computation---such as window functions, angular power
spectra, or multipoles---can operate on either without requiring
modification. This structural approach, rather than relying on
inheritance hierarchies, enables the seamless integration of external
codes without altering their source. As a result, it provides a clear
pathway for the community to connect their own tools, provided they
comply with the protocol.

\hypertarget{automatic-differentiation-with-jax}{%
\subsection{Automatic Differentiation with
JAX}\label{automatic-differentiation-with-jax}}

One of \texttt{cloelib}'s unique features is its support for automatic
differentiation:

\begin{lstlisting}[language=python]
import jax
import jax.numpy as jnp
from cloelib.cosmology.jax_cosmology import JAXBackground

def compute_observable(params):
    """Compute the angular diameter distance at z=1 from (H0, Omega_m)."""
    H0, Omega_m = params
    bg = JAXBackground(
        H0=H0, Omega_cdm0=Omega_m - 0.05, Omega_b0=0.05,
        As=2e-9, ns=0.96, w0=-1.0, wa=0.0, Omega_k0=0.0,
        mnu=0.0, N_mnu=0, gamma_MG=0.545,
    )
    return bg.angular_diameter_distance(jnp.array([1.0]))[0]

# Compute gradients with automatic differentiation
grad_fn = jax.grad(compute_observable)
gradients = grad_fn(jnp.array([70.0, 0.3]))
# gradients[0] = dD_A/dH0,  gradients[1] = dD_A/dOmega_m
\end{lstlisting}

The \texttt{JAXBackground} and
\texttt{JAXLinearPerturbations}/\texttt{JAXNonLinearPerturbations}
classes are fully \texttt{jit}-compilable and differentiable through
\texttt{jax.grad}, \texttt{jax.jacobian}, and \texttt{jax.hessian}. This
enables Hamiltonian Monte Carlo samplers, variational inference, and the
training of neural-network emulators whose inputs are cosmological
parameters.


\hypertarget{computational-times}{%
\subsection{Computational times}\label{computational-times}}

\texttt{cloelib} exhibits performance comparable to other tools
available in the community, despite being implemented exclusively in
Python. The computational cost of the \texttt{Background}-compatible
classes is negligible, 
whereas the runtime
of the \texttt{Perturbation}-compatible classes depends on the choice of
backend, namely whether a Boltzmann solver or an emulator is employed.

Below, we provide representative estimates of the computational time
required to evaluate key cosmological
observables\footnote{Apple MacBook Pro (Model Mac16,1) with Apple M4 chip (10 cores: 4 performance, 6 efficiency), 16 GB RAM, running macOS.}.
These comprise photometric probes---cosmic shear, photometric galaxy
clustering, and galaxy--galaxy lensing, the so-called 3x2-pt
analysis---calculated in harmonic space (angular power spectra). We also
present corresponding estimates for full-shape analyses of spectroscopic
galaxy clustering, in Fourier space (Legendre multipoles), utilising
\texttt{comet-emu} as the backend. For all cases, \texttt{CAMB} is
employed to compute \texttt{Background} quantities, while for
photometric probes, \texttt{HMCode2020emu} is used as the
\texttt{Perturbations} backend.

For photometric analyses, initialising the shear and position tracer
classes---each conforming to the \texttt{Tracer} protocol---typically
requires approximately 0.4 seconds to compute the lensing efficiency
(utilised for both shear and the magnification systematic effect in the
position tracer). This quantity is cached, ensuring that all subsequent
calls to either tracer are effectively instantaneous.

The computation of angular power spectra for a 3x2-pt analysis---using
3000 multipole values, 512 $k$-values, 500 redshift z-values, and linear
galaxy and magnification bias---with six redshift bins (78 spectra in
total), using Limber approximation, takes approximately 0.06 seconds per
call. This follows a one-second initialization phase, during which
\texttt{jit}-compiled quantities are cached to accelerate subsequent
integral evaluations. For a \emph{Euclid}-like final data release
configuration as described in \citet{Euclid:2024}, with thirteen redshift bins, the computation of 351 angular power spectra for a 3x2-pt analysis requires approximately 0.1 seconds per call after initialization and caching. This is an important speed up
achievement with respect to the former \texttt{CLOE} software.

For full-shape spectroscopic Legendre multipoles, the computation takes
approximately 3 milliseconds for a single redshift bin.

\hypertarget{performance-profiling}{%
\subsection{Performance Profiling}\label{performance-profiling}}

\texttt{cloelib} includes function-level profiling via the
\texttt{@profile\_function} decorator, configurable sampling (default
interval: 0.001 s) to balance overhead and granularity, and timestamped
interactive HTML reports for run-to-run comparison. Profiling can be
controlled through environment variables or function calls (including
enable/disable and output configuration), and it includes safeguards to
avoid redundant profiling in nested decorated calls.

\begin{lstlisting}[language=python]
from cloelib.profiling import (
  profile_function, 
  enable_profiling, 
  disable_profiling
)

from cloelib.summary_statistics.angular_correlation_function_wigner import (
    AngularCorrelationFunctionWigner
)

# Define at which angular separation to evaluate the correlation function
theta_rad = np.deg2rad(np.geomspace(5, 100, 20)/60)  
ells_integration = np.arange(2, 60000)

# Initialize correlation function calculation object
correlation_function_GG = AngularCorrelationFunctionWigner(twopoint_pospos, ells_integration, hmcode2020emu_nonlinear.k)

#apply the wrapper
get_xi_profiled = profile_function(correlation_function_GG.get_xi)

# Enable the profiling
enable_profiling()

xi_pospos = get_xi_profiled(theta_rad)

# Disable the profiling
disable_profiling()
\end{lstlisting}


\hypertarget{documentation}{%
\section{Documentation}\label{documentation}}

Comprehensive documentation for \texttt{cloelib} is available at
\href{https://cloe-org.github.io/cloelib/dev/home/}{cloe-org.github.io/cloelib/dev/home/}.
The documentation includes detailed API references, installation
instructions, explanations about the software structure, and guides for
integrating cloelib into your analysis workflows.

For practical examples, example scripts, and interactive tutorials,
visit the
\href{https://github.com/cloe-org/playground}{cloe-org/playground}
repository, which hosts a collection of Jupyter notebooks showcasing
typical use cases and advanced features.

\texttt{cloelib} output formats for both photometric and spectroscopic
observables are compliant with
\texttt{euclidlib}\footnote{\href{https://euclidlib.readthedocs.io/en/latest/}{https://euclidlib.readthedocs.io/en/latest/}}
formats, using
\texttt{cosmolib}\footnote{\href{https://github.com/astro-ph/cosmolib}{https://github.com/astro-ph/cosmolib}}
dataclasses.

\section{Availability}
\noindent
\textbf{Source:} \href{https://github.com/cloe-org/cloelib}{github.com/cloe-org/cloelib} \\
\textbf{License:} MIT. \\
\textbf{Install (PyPI):} \texttt{pip\,\, install\,\, cloelib} \\
\textbf{Documentation:} \href{ https://your_package.readthedocs.com}{https://cloe-org.github.io/cloelib/dev/home/}\\
\textbf{\texttt{conda/mamba} environments:} \href{https://github.com/cloe-org/cloe-org-environments}{github.com/cloe-org/cloe-org-environments}\\
\textbf{Examples:}
\href{https://github.com/cloe-org/playground}{github.com/cloe-org/playground}\\


\section*{Acknowledgments}

We thank the broader CLOE software development team for foundational work that motivated this
library. We thank Fabrice Roy for helping deplying the docs. The scientific development of cloe-org is coordinated through the Joint Cosmology Key Project one (DR1-KP-JC-1) of the Euclid Consortium, led by C. Carbone, M. Crocce, and I. Tutusaus. In this context, \texttt{cloe-org} has been adopted as the default analysis code of the Euclid Consortium.  G.C.H. acknowledges that this project is part of the project
UNICORN with file number VI.Veni.242.110 of the research programme
Talent Programme Veni Science domain 2024 which is (partly) financed by
the Dutch Research Council (NWO) under the grant
https://doi.org/10.61686/ZCPQI32997. M.B. acknowledges support from the
Natural Sciences and Engineering Research Council of Canada (NSERC).
C.M. is supported by the Agenzia Spaziale Italiana project ``Attività
scientifica per la missione Euclid -- fase E ACCORDO ATTUATIVO
n.~2024-10-HH.0.'' B.B. is supported by a UK Research and Innovation
Stephen Hawking Fellowship (EP/W005654/2). E.B. and K.T. acknowledge
support by the European Union's Horizon Europe research and innovation
program under the Marie Sklodowska-Curie COFUND Postdoctoral Programme
grant agreement No.101081355- SMASH and from the Republic of Slovenia
and the European Union from the European Regional Development Fund. A.H.
acknowledges the support of a Royal Society University Research
Fellowship. A.H.W. is supported by the Deutsches Zentrum für Luft- und
Raumfahrt (DLR), under project 50QE2305, made possible by the
Bundesministerium für Wirtschaft und Klimaschutz, and acknowledges
funding from the German Science Foundation DFG, via the Collaborative
Research Center SFB1491 ``Cosmic Interacting Matters - From Source to
Signal.'' R.K. is supported by UK STFC grant ST/X001040/1. N.G. acknowledges the support of the Royal Society as a Newton International Fellow (NIF\textbackslash R1\textbackslash 252792) and by the STFC (ST/B001175/1). I.T. acknowledges support form the Spanish Ministerio de Ciencia, Innovaci\'on y Universidades, projects PID2022-141079NB, PID2022-138896NB; the European Research Executive Agency HORIZON-MSCA-2021-SE-01 Research and Innovation programme under the Marie Sk\l odowska-Curie grant agreement number 101086388 (LACEGAL) and the programme Unidad de Excelencia Mar\'{\i}a de Maeztu, project CEX2020-001058-M.  

We acknowledge EuroHPC Joint Undertaking for awarding the project ID
EHPC-EXT-2024E02-083 access to Leonardo hosted by CINECA, Italy. We
acknowledge the use of Spanish Supercomputing Network (RES) resources
provided by the Barcelona Supercomputing Center (BSC) in MareNostrum 5
under allocations AECT-2024-3-0020, 2025-1-0045, 2025-2-0046,
2025-3-0036. We acknowledge support from the European Research Council
(ERC) under the European Union's Horizon 2020 research and innovation
program with Grant agreement No.~101053992 for computational resources.
We acknowledge the use of computing resources at the JURECA cluster of the
Forschungszentrum Jülich (FZJ) under the project name paj2526.

\AckEC


\section*{Author Contributions}

In accordance with JOSS guidelines, we describe individual contributions
below. Authors are listed in alphabetical order. All Tier 1 authors are
core maintainers and original developers of the \textbf{cloe-org} organisation, responsible for
the long-term sustainability of \texttt{cloelib}, the review of pull
requests, and leadership of technical discussions.

\begin{itemize}
\tightlist
\item
  \textbf{M. Bonici}: Core architecture and protocol design;
  implementation of the JAX cosmology backends; lensing tracer kernels
  (including massive neutrino contributions); correlation function
  module and performance optimisation; caching system with JAX
  \texttt{lax} conditional compatibility; licence and project
  governance.
\item
  \textbf{G. Cañas-Herrera}: Project overview and release management;
  core architecture of the software, protocol and class inheritance
  design; continuous integration (CI) pipeline configuration; pre-commit and code-quality
  tooling; issue and pull-request templates; README, documentation, and
  community contribution tracking (\texttt{all-contributors});
  \texttt{pyproject.toml} versioning and release workflows. Developed models for
  systematic shear and calibration of photometric redshift nuisance
  parameters. Homogenisation of output format for observables. Resources, Writing - Original Draft, Visualization, Funding.
\item
  \textbf{P. Carrilho}: Photometric observable module linear galaxy bias
  models with JAX-compatible conditional logic; HMCode2020Emu baryonic
  feedback support and further extrapolation support; \texttt{CAMB}
  dark-energy model configuration (PPF); mixing-matrix and
  pseudo-\(C_\ell\) corrections; \texttt{interpax}-based interpolation
  in the extrapolator; growth-rate and matter power spectrum
  redshift/scale interfaces, implementation of EuclidEmulator2.
\item
  \textbf{S. Casas}: Implementation of the \texttt{CLASS} cosmology
  backend and its integration with the \texttt{Background} and
  \texttt{Perturbations} protocols; fixes to transverse-distance
  computations across \texttt{CAMB}, \texttt{CLASS}, and JAX backends;
  cosmology protocol refinements; CI pipeline and dependency updates.
\item
  \textbf{C. Moretti}: Spectroscopic analysis infrastructure: PBJ
  interface and RSD power spectrum fixes; BAO \(\alpha\)-parameter
  module and Alcock--Paczynski distortion utilities; extraction of
  \(r_\mathrm{drag}\) from the background for BAO analyses; Legendre
  multipole summary statistics; version management and repository
  clean-up of deprecated directories.
\item
  \textbf{A. Pezzotta}: Spectroscopic analysis infrastructure:
  \texttt{comet-emu} interface and EFT and VDG spectroscopic power
  spectrum implementations; survey window-function convolution of power
  spectrum building blocks; Legendre multipole computation optimisation
  (\texttt{np.einsum}); documentation of the spectroscopic observable
  interface.
\end{itemize}

C. Carbone, M. Crocce, and I. Tutusaus: Project administration, Supervision, Validation. The contributions of all remaining authors have been tracked using the \href{https://github.com/all-contributors/all-contributorsDESI_review}{all-contributors}
bot, following the specification of the same name. A full, categorised
breakdown of each contributor's role---including code, documentation,
testing, ideas, project management, and more---is available in the
\texttt{README} of the \texttt{cloelib} repository, fully detailed
within the \texttt{cloelib} docs.

\bibliographystyle{mnras}
\bibliography{main_joss}

\end{document}